\newcommand{\msbar}{{\overline {\rm MS}}}

\documentclass{PoS}

\title{Heavy-light mesons in 2+1 flavor lattice QCD}

\ShortTitle{Heavy-light mesons in 2+1 flavor lattice QCD}

\author{
 \speaker{Y. Namekawa} \thanks{E-mail: namekawa@ccs.tsukuba.ac.jp}
 for PACS-CS Collaboration,
 \\
 \llap{}Center for Computational Sciences, University of Tsukuba, Tsukuba,
 Ibaraki 305-8577, Japan
}

\abstract{
Heavy-light meson system is
investigated using the relativistic heavy quark action 
on the 2+1 dynamical flavor PACS-CS configurations at the lattice spacing $a^{-1}=2.2$ GeV
and the spatial extent $L=3$ fm. 
Dynamical up-down and strange quark masses as well as
the valence charm quark mass are set around
their physical values.
We measure the charm-$ud$ and charm-strange
meson masses and decay constants.
Our results are consistent with the experimental values
except the hyperfine splitting of the charm-strange meson.
We also estimate the CKM matrix elements in the second row.
          }

\FullConference{The XXVII International Symposium on Lattice Field Theory - LAT2009\\
		 July 26-31 2009\\
		 Peking University, Beijing, China}

\begin{document}

\section{Introduction}
\label{section:introduction}

Precise determination of the CKM matrix is an important step to establish the 
validity range of the Standard Model, thereby setting the basis for exploring 
physics at smaller space-time scales.
In this respect the CKM matrix elements in the second row,
especially $|V_{cd}|$ and $|V_{cs}|$, 
are still poorly determined with errors in the 10\% range~\cite{PDG_2008}.
This situation contrasts sharply with those of the first row
where $|V_{ud}|$ and $|V_{us}|$ are known at sub percent level.  

Lattice QCD provides a variety of means for precise determination of the second row of 
the CKM matrix.  With $|V_{cs}|$, for example, the leptonic decay width of the $D_s$ meson 
$\Gamma(D_s \rightarrow l \nu)$ is given by 
\begin{eqnarray}
 \Gamma(D_s \rightarrow l \nu)
 = \frac{G_F^2}{8 \pi} m_l^2 m_{D_s} f_{D_s}^2 
   \left( 1 - \frac{m_l^2}{m_{D_s}^2} \right)^2 |V_{cs}|^2.
\end{eqnarray}
so that a lattice determination of the decay constant $f_{D_s}$ combined with the 
experimental value of $\Gamma(D_s \rightarrow l \nu)$ will allow us to extract $|V_{cs}|$.
The other matrix element $|V_{cd}|$ can be obtained in the same way.

An intriguing topic in the charm quark sector
is a disagreement of the value of the $D_s$ meson decay constant
from lattice QCD and experiment.
In 2007 HPQCD and UKQCD Collaboration reported a calculation of this quantity 
using the HISQ action for heavy quark on the 2+1 dynamical flavor MILC configurations~\cite{HPQCD}. 
Their value $f_{D_s}=241\pm 3$~MeV exhibited significant deviation from 
those of the CLEO and Belle experiments at the time. 
The latest CLEO value~\cite{CLEO}  $f_{D_s}=259.5\pm 7.3$~MeV, though smaller than 2007, 
is still two standard deviations away. 

FNAL Collaboration also calculated the decay constant on the MILC configurations 
using the clover heavy quark formalism.  Their original value $f_{D_s}=249(3)(16)$~MeV 
\cite{FNAL_2005} was also smaller than experiment, but an update this year \cite{FNAL_2009} 
including a 2\% change in the scale setting of the MILC ensemble moved the value 
up to $f_{D_s}=260(10)$~MeV.

In this paper we present a status report of our work on the charmed mesons using the 
relativistic heavy quark formalism on the 2+1 dynamical 
flavor PACS-CS configurations.

\section{Set up}
\label{section:setup}

The charm quark system is simulated with a relativistic
heavy quark action on the 2+1 flavor lattice QCD configurations.
The configurations are generated by the PACS-CS Collaboration~\cite{PACS_CS}
using the nonperturbatively $O(a)$-improved Wilson quark action 
with $c_{\rm SW}^{\rm NP}=1.715$~\cite{Csw_NP} and the Iwasaki gauge action.
The lattice size is $32^3\times 64$ whose spatial extent is $L=3$ fm
with the lattice spacing of $a=0.09$ fm. The dynamical up-down quark masses
range from 8 MeV down to 3 MeV, which is close to the the physical value.
We utilize the same quark action for the light quark propagators in our calculation.

For the heavy quarks, we employ a relativistic heavy quark action proposed in Ref.~\cite{AKT}.
The cutoff errors are reduced
from $O((m_Q a)^n)$
to $O(f(m_Q a)(a \Lambda_{QCD})^2)$
where $f(m_Q a)$ is an analytic function around $m_Q a = 0$.
We use one-loop perturbative values for parameters in the heavy quark action~\cite{AKT_parameters}.
In addition, nonperturbative contributions at the massless limit is included for the clover coefficients,
and the parameter $\nu$ is adjusted nonperturbatively
to reproduce the relativistic dispersion relation of the charmonium.
Our valence heavy quark masses are always tuned
to the physical charm quark mass point,
which is determined
by the spin-averaged $1S$ state of the charmonium,
$M(1S) = (M_{\eta_c} + 3 M_{J/\psi})/4 = 3.0677(3)$~GeV.
Our results for charmonium spectrum
have been reported in Ref.~\cite{lattice2008}.

Table~\ref{table:statistics} summarizes our simulation parameters and the
statistics of the configuration sets we have used for the heavy quark 
measurements.
The number of the source points is quadrupled to suppress statistical fluctuations.

\begin{table}[t]
\caption{Simulation parameters. 
Quark masses are perturbatively renormalized
in the $\msbar$ scheme. 
The renormalization scale is $\mu=1/a$ for each simulation point
and $\mu=2$ GeV for the physical point.}
\label{table:statistics}
\begin{center}
\begin{tabular}{cccccc}
\hline
\multicolumn{1}{c}{$\kappa_{\rm ud}$} & \multicolumn{1}{c}{$\kappa_{\rm s}$} &
 \multicolumn{1}{c}{$m_{\rm ud}^{\msbar}(\mu)$} [MeV] & 
 \multicolumn{1}{c}{$m_{\rm s}^{\msbar}(\mu)$}  [MeV] & 
 \multicolumn{2}{c}{\#conf} \\
& & &  &
 measured & MD time/total
\\ \hline
 0.13781          & 0.13640 &
 3.5(2) & 86.4(1) &
  65    & 1625/1625
 \\
 0.13770          & 0.13660 &
 8.3(5) & 74.1(4) &
  60    & 1500/1500
 \\
 0.137785         & 0.13660 &
 3.5(1) & 72.8(2) &
 200    & 1000/2000
\\ \hline
\multicolumn{2}{c}{physical point} & 2.53(5) & 72.7(8) & &
\\ \hline
\end{tabular}
\end{center}
\end{table}

\begin{figure}[t]
\begin{center}
 \includegraphics[width=75mm]{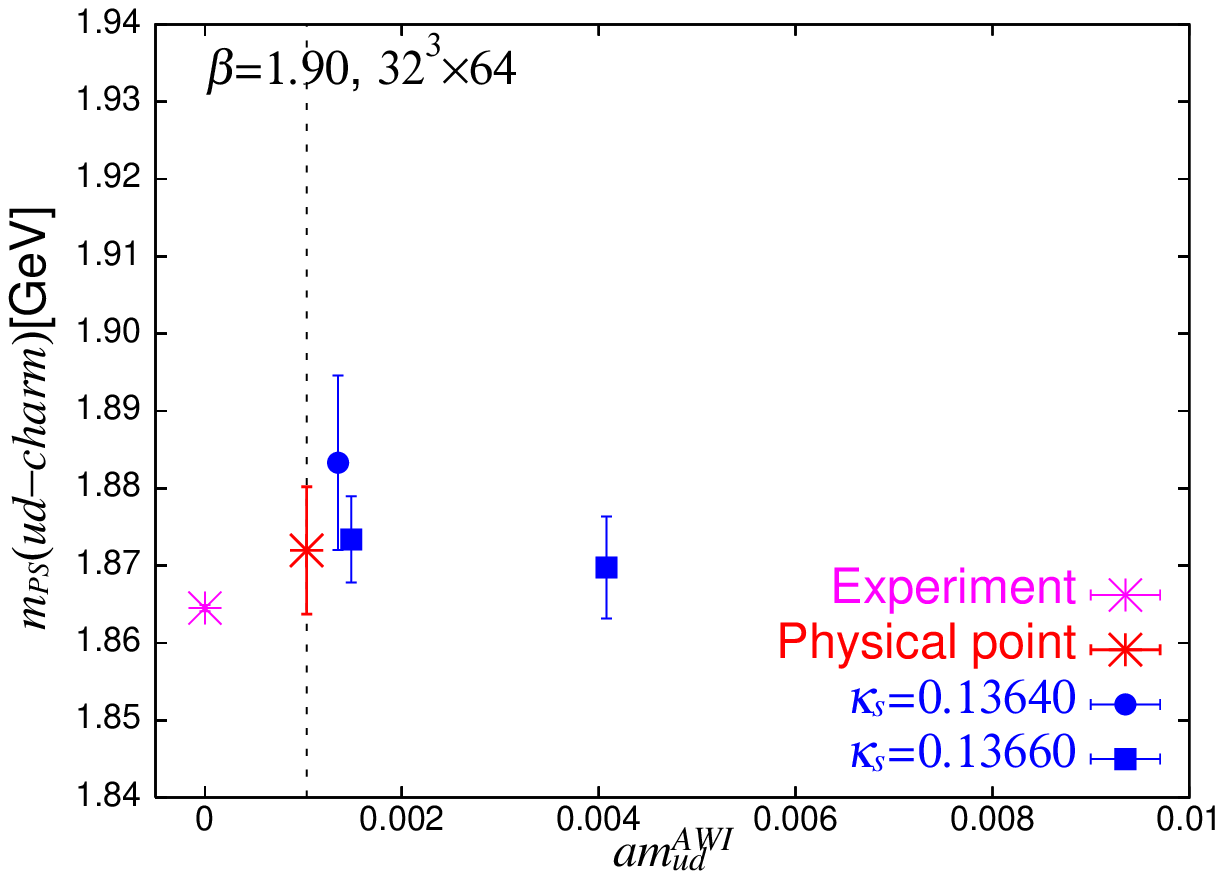}
 \includegraphics[width=75mm]{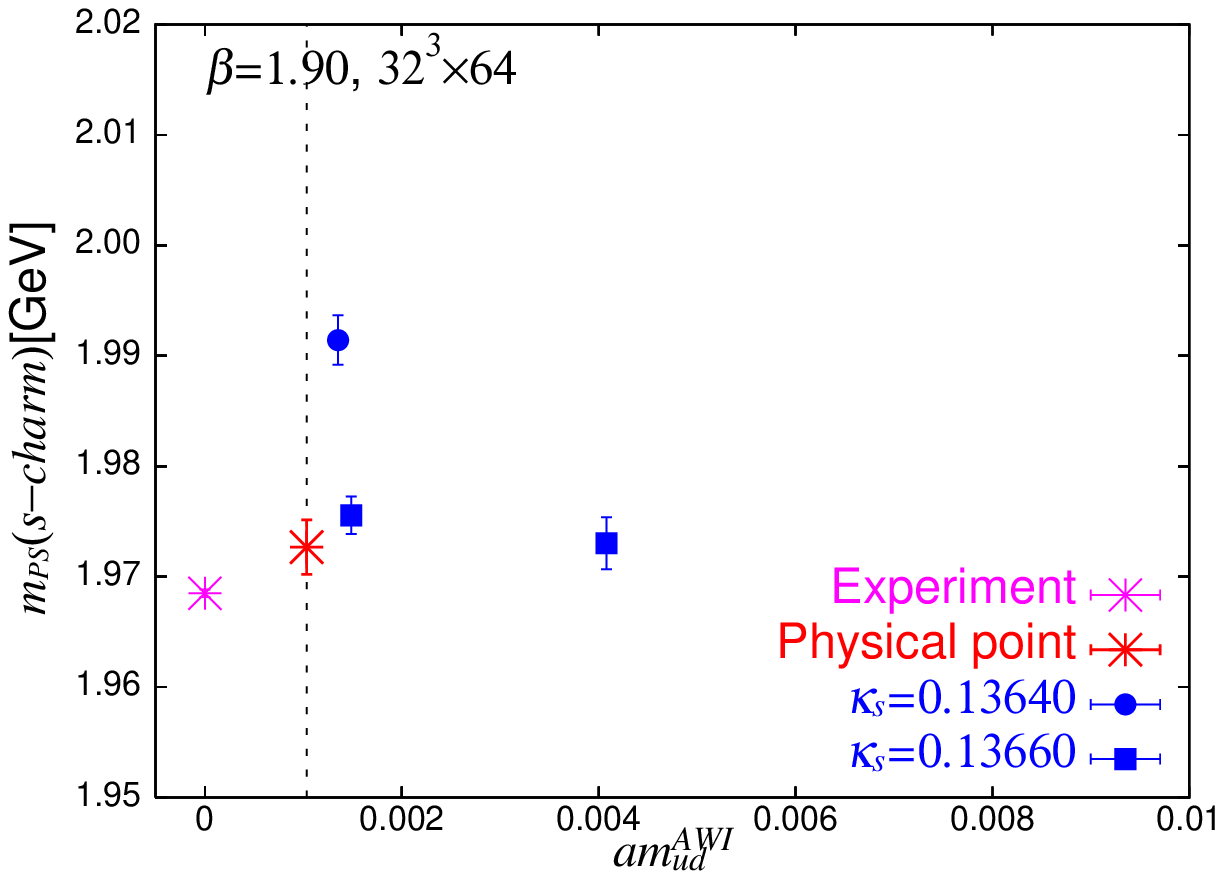}
 \caption{
  $m_{\rm ud}^{sea}$ quark mass dependence of $m_D$(left panel)
  and $m_{D_s}$(right panel).
 }
 \label{figure:masses_1}
\end{center}
\end{figure}

\begin{figure}[t]
\begin{center}
 \includegraphics[width=75mm]{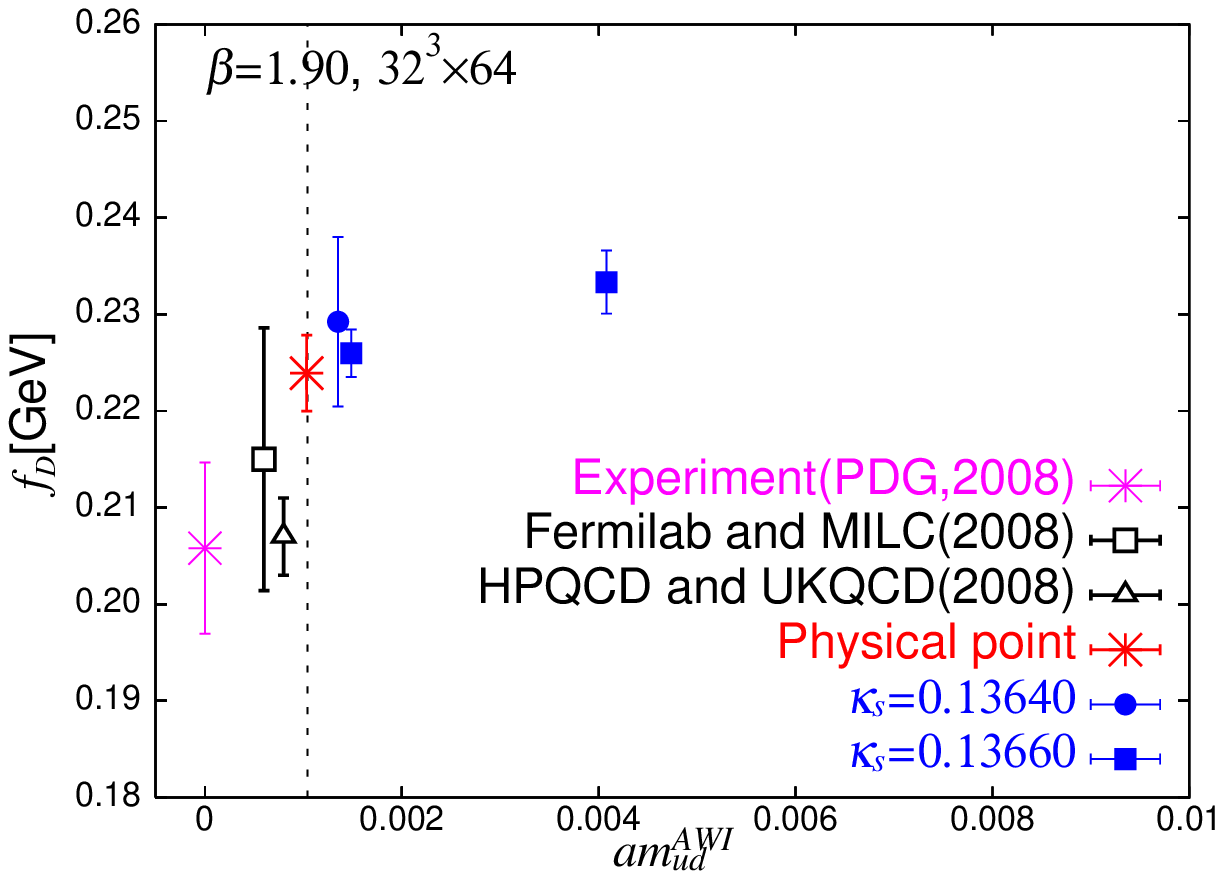}
 \includegraphics[width=75mm]{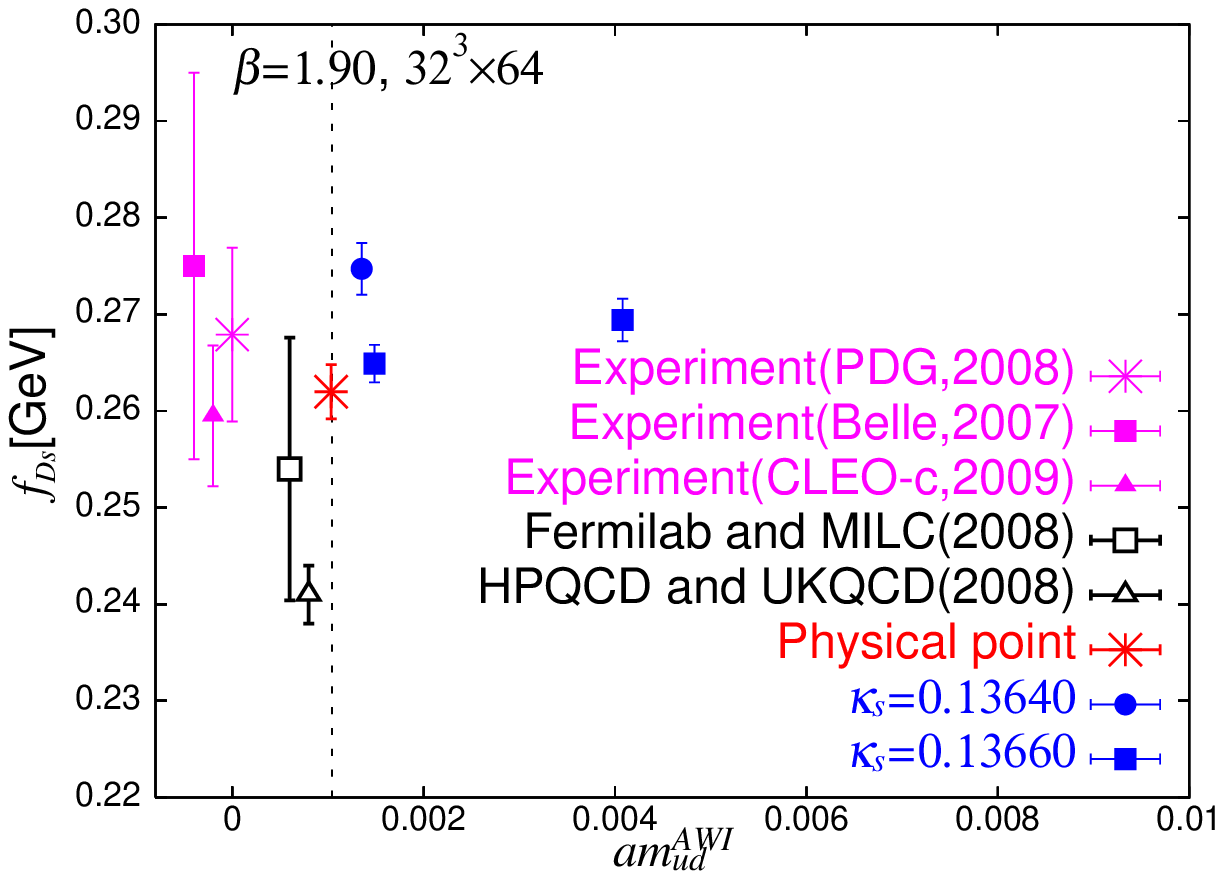}
 \caption{
  $m_{\rm ud}^{sea}$ quark mass dependence of $f_D$(left panel)
  and $f_{D_s}$(right panel).
 }
 \label{figure:decay_constants}
\end{center}
\end{figure}

\begin{figure}[t]
\begin{center}
 \includegraphics[width=75mm]{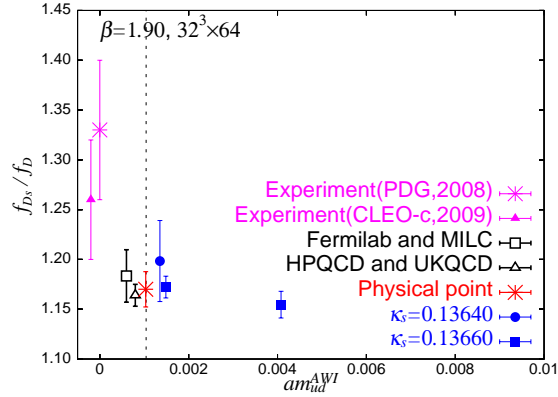}
 \caption{
  $m_{\rm ud}^{sea}$ quark mass dependence of $f_{D_s}$ over $f_D$.
 }
 \label{figure:f_D_s_over_f_D}
\end{center}
\end{figure}

\section{Results for $D$ and $D_s$ meson masses and decay constants}
\label{section:result_1}

We present our results for $D$ and $D_s$ meson masses
and their decay constants.
Since the physical charm quark mass has already been fixed with the heavy-heavy spectrum,
all heavy-light meson quantities can be predicted.

Figure~\ref{figure:masses_1} compares our results for 
the $D$ and $D_s$ meson masses with the experimental values~\cite{CLEO}.
Since our sea quark masses are close to the physical values,
we extrapolate our results with a linear function
of the up-down and the strange quark masses to the physical point,
\begin{eqnarray}
 M = A + B (m_{\rm ud} - m_{\rm ud}^{\rm phys})
       + C (m_{\rm s}  - m_{\rm s}^{\rm phys}).
\end{eqnarray}
Our results at the physical point (red bursts) reproduce the experimental spectrum well.

We also calculate the decay constants of heavy-light pseudoscalar mesons,
\begin{eqnarray}
 < 0 | A_{\mu}^{imp} | PS(p) >
 &=& i f_{PS} p_{\mu},\\
 A_{\mu}^{imp}
 &=& \sqrt{2 \kappa_q} \sqrt{2 \kappa_Q} Z_{A_{\mu}}
     \left\{ \bar{q}(x) \gamma_{\mu} \gamma_{5} Q(x)
     \right. \\
 &&  \left. -c_{A_{\mu}}^{+} \partial_{\mu}^{+} \left( \bar{q}(x) \gamma_{\mu} \gamma_{5} Q(x) \right)
            -c_{A_{\mu}}^{-} \partial_{\mu}^{-} \left( \bar{q}(x) \gamma_{\mu} \gamma_{5} Q(x) \right)
     \right\}.
\end{eqnarray}
For the renormalization factor and the improvement coefficients of the axial current, 
we employ one-loop perturbative values~\cite{Z_factors}.
Furthermore, the nonperturbative contribution at the massless limit
is incorporated to the improvement coefficient $c_{A_4}^{+}$ by writing 
\begin{eqnarray}
 c_{A_4}^{+}=(c_{A_4}^{+}(m_Q a) - c_{A_4}^{+}(0))^{\rm PT} + c_{A}^{\rm NP}
\end{eqnarray}
with $c_{A}^{\rm NP} = -0.03876106$~\cite{NP-c_A}.

Figure ~\ref{figure:decay_constants} shows the decay constants,
where we also plot other 2+1 flavor lattice QCD results
from HPQCD and UKQCD Collaboration~\cite{HPQCD} and FNAL group~\cite{FNAL_2009}, 
as well as the experimental values~\cite{PDG_2008,CLEO,CLEO2}, for comparison.
Our value for $f_{D_s}$ agrees with the experimental determinations, while that for $f_D$ is 
somewhat larger. 
Comparing the three sets of lattice determinations, we observe, both for $f_D$ and $f_{D_s}$,  
an agreement between our values and those of the FNAL group, while there seems to be a clear 
discrepancy between our values and that by the HPQCD and UKQCD Collaboration.
Possible reasons are our use of perturbative renormalization factors and the necessity of taking the 
continuum extrapolation.

In Fig.~\ref{figure:f_D_s_over_f_D} we plot
the ratios of $f_{D_s}$ to $f_D$
in which uncertainties coming from the perturbative renormalization factors cancel out, 
and perhaps that of the lattice cutoff to some extent.
It is interesting that all three lattice results are mutually consistent with small errors of a 
few percent, but that they seem to lie at the lower edge of the experimental value.

\section{Results of hyperfine splittings for charm-$ud$ and charm-strange mesons}
\label{section:result_2}

We evaluate the hyperfine splittings for charm-$ud$ and charm-strange mesons.
Masses of $D^*$ and $D_s^*$ are obtained from their two-point functions.
$D^*$ and $D_s^*$ decays are prohibited energetically on our lattice.

\begin{figure}[t]
\begin{center}
 \includegraphics[width=75mm]{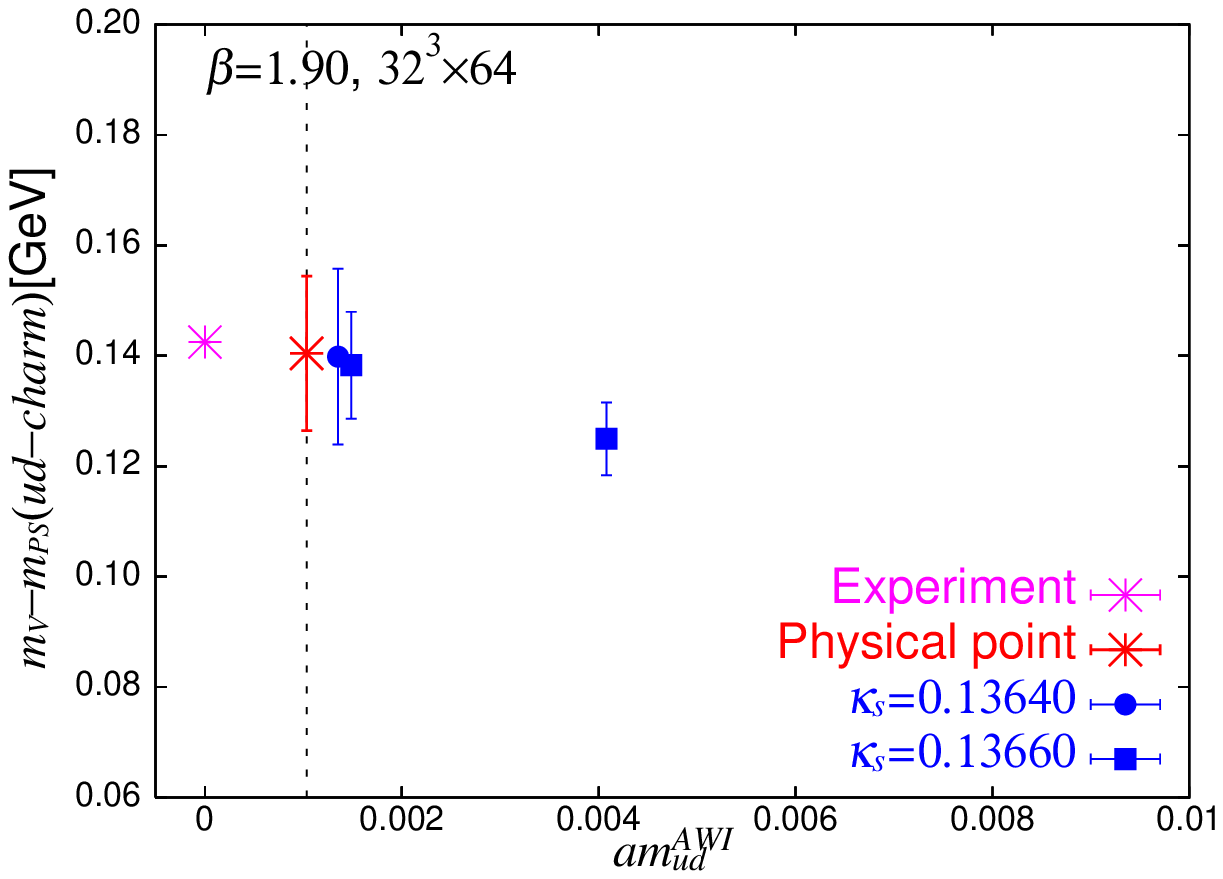}
 \includegraphics[width=75mm]{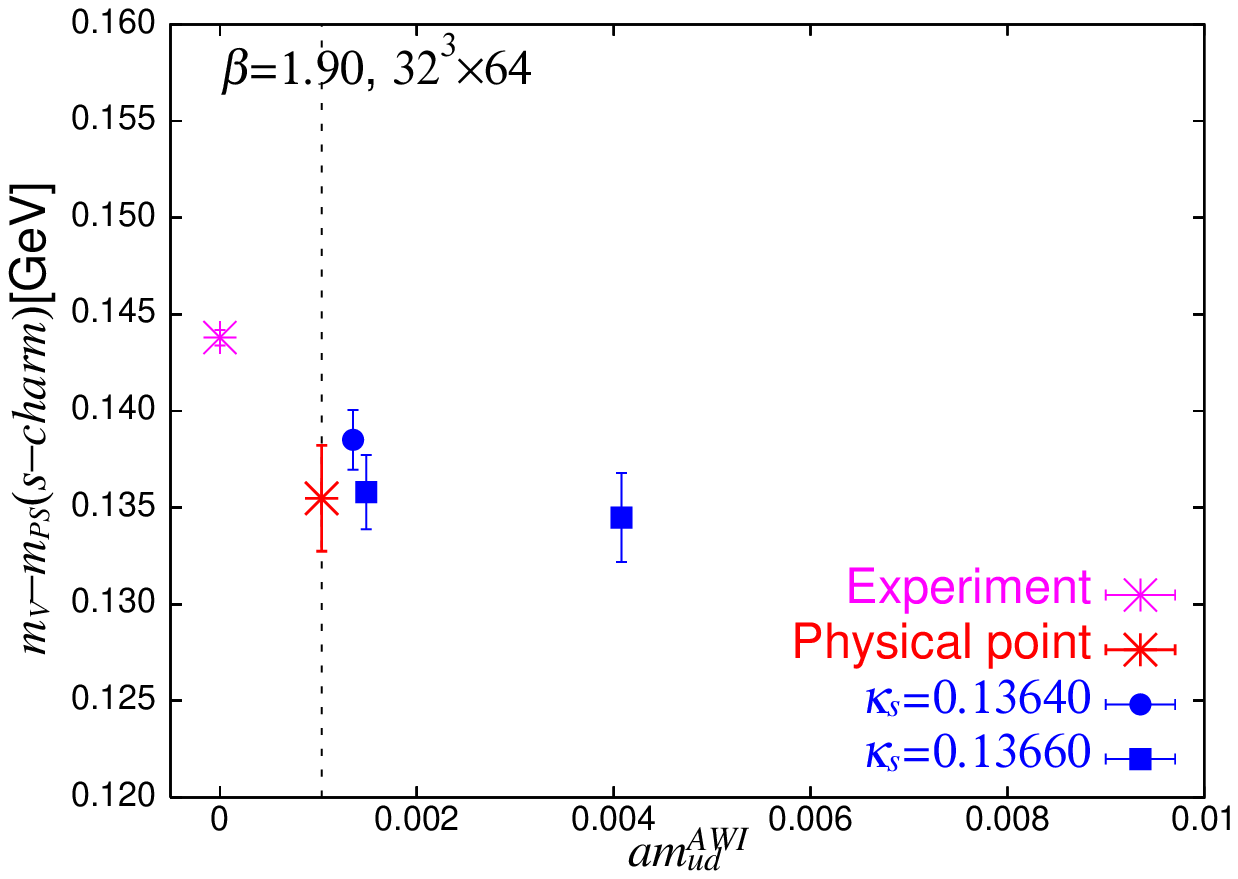}
 \caption{
  $m_{\rm ud}^{sea}$ quark mass dependence of hyperfine splittings for charm-$ud$(left panel)
  and and charm-strange mesons (right panel).
 }
 \label{figure:masses_2}
\end{center}
\end{figure}

Figure~\ref{figure:masses_2} represents our results for 
the hyperfine splittings with the experimental values~\cite{PDG_2008}.
For $D^*$--$D$ mass difference,
our result at the physical point (red bursts) reproduces the experimental spectrum.
On the other hand,
$D^*_s$--$D_s$ mass difference shows
10\% deviation from the experimental value.
A possible origin of this discrepancy is the scaling violation.

\section{Estimating the CKM matrix elements}
\label{section:CKM}

We attempt to estimate the CKM matrix elements from our data of
$D$ and $D_s$ meson masses and decay constants combined with experimental values of
leptonic decay widths.
Using the CLEO value of $\Gamma(D_s \rightarrow l \nu)$~\cite{CLEO},
we find 
\begin{eqnarray}
 |V_{cs}|(lattice) &=& 0.98(2)(3)+ O(g^2 a),\\
 |V_{cs}|(PDG)     &=& 1.04(6)~\cite{PDG_2008} 
\end{eqnarray}
The first error is statistical,
and the second is experimental.
For comparison, the PDG value is listed.
Our preliminary result is consistent with the PDG value.
We note that our estimate of $|V_{cs}|$ still has
a discretization error of $O(g^2 a)$ due to the use of perturbative renormalization 
factor.
We must take the continuum extrapolation to remove this source of uncertainty.

We can estimate $|V_{cd}|$ from the $D$ meson mass and decay constant
and CLEO value of $\Gamma(D \rightarrow l \nu)$~\cite{CLEO2}:
\begin{eqnarray}
 |V_{cd}|(lattice) &=& 0.207(2)(9) + O(g^2 a),\\
 |V_{cd}|(PDG)     &=& 0.230(11)~\cite{PDG_2008} 
\end{eqnarray}
Our estimate is smaller than the PDG value by about 10\%.
For completeness we also record the ratio $|V_{cs}| / |V_{cd}|$ for which 
$O(g^2a)$ errors drop out, and is replaced with scaling violation of $O(a^2)$:
\begin{eqnarray}
 \frac{|V_{cs}|}{|V_{cd}|}(lattice) &=& 4.72(13)(26) + O(a^2),\\
 \frac{|V_{cs}|}{|V_{cd}|}(PDG)     &=& 4.52(34)~\cite{PDG_2008}.
\end{eqnarray}

\section{Conclusion}
\label{section:conclusion}

We measured charm-$ud$ and charm-strange
meson masses and decay constants
with the relativistic heavy quark action 
on the 2+1 dynamical flavor PACS-CS configurations
at $a^{-1}=2.2$ GeV.
Since our sea quark masses are close to the physical point,
sea quark mass corrections in our results are small
and only short extrapolations to the physical point are needed.

We found our results are consistent with the experimental values
in two standard deviations.
We do not observe discrepancy from experiments in $f_{D_s}$.

Combining our data of masses and decay constants
with experimental values of leptonic decay widths,
CKM matrix elements are evaluated.
Our estimates still have discretization errors,
which is expected to be 1\% by the order counting.
Continuum extrapolation is needed
to achieve a few percent accuracy.

\begin{acknowledgments}
Numerical calculations for the present work have been carried out
on the PACS-CS computer
under the ``Interdisciplinary Computational Science Program'' of
Center for Computational Sciences, University of Tsukuba.
This work is supported in part by Grants-in-Aid for Scientific Research
from the Ministry of Education, Culture, Sports, Science and Technology
(Nos.
16740147,
17340066,
18104005,
18540250,
18740130,
19740134,
20105002,
20340047,
20540248,
20740123,
20740139
).
\end{acknowledgments}



\begin{thebibliography}{99}
 \bibitem{PDG_2008}
 Particle Data Group, C. Amsler {\it et al},
 {\em Phys. Lett. B} {\bf 667} (2008) 1.
%
 \bibitem{HPQCD}
 HPQCD and UKQCD Collaboration, E.~Follana {\it et al.},
 {\em Phys. Rev. Lett.} {\bf 100} (2008) 062002.
%
 \bibitem{CLEO}
 CLEO Collaboration., J.P.~Alexander {\it et al.},
 {\em Phys. Rev. D} {\bf 79} (2009) 052001.
%
 \bibitem{FNAL_2005}
 C. Aubin {\it et al}, {\em Phys. Rev. Lett.} {\bf 95} (2005) 122002.
%
 \bibitem{FNAL_2009}
 Fermilab Lattice and MILC Collaboration, A.~Bazavov {\it et al},
 \pos{PoS(LATTICE 2009)249}.
%
 \bibitem{PACS_CS}
 PACS-CS Collaboration, S.~Aoki {\it et al.},
 {\em Phys. Rev. D} {\bf 79} (2009) 034503;\\
 PACS-CS Collaboration, Y.~Kuramashi {\it et al.},
 \pos{PoS(LATTICE 2009)110}.
%
 \bibitem{Csw_NP}
 CP-PACS and JLQCD Collaborations, S. Aoki {\it et al.},
 {\em Phys. Rev.} {\bf D73}, 034501 (2006).
%
 \bibitem{AKT}
 S.~Aoki, Y.~Kuramashi and S.~Tominaga,
 {\em Prog. Theor. Phys.} {\bf 109} (2003) 383.
%
 \bibitem{AKT_parameters}
 S.~Aoki, Y.~Kayaba and Y.~Kuramashi,
 {\em Nucl. Phys. B} {\bf 697} (2004) 271.
%
 \bibitem{lattice2008}
 PACS-CS Collaboration, Y.~Namekawa {\it et al.},
 \pos{PoS(LATTICE 2008)121}.
%
 \bibitem{Z_factors}
 S.~Aoki, Y.~Kayaba and Y.~Kuramashi,
{\em Nucl. Phys. B} {\bf 689} (2004) 127.
%
 \bibitem{NP-c_A}
 CP-PACS/JLQCD and ALPHA Collaboration, T.~Kaneko {\it et al.},
 {\em JHEP} {\bf 0704} (2007) 092.
%
 \bibitem{CLEO2}
 CLEO Collab., B.I.Eisenstein {\it et al.},
 {\em Phys. Rev. D} {\bf 78} (2008) 052003.
\end{thebibliography}
\end{document}